# Photo-transmutation of long-lived radionuclide $^{135}$Cs by laser-plasma driven electron source

## X.L. WANG,[1] W. LUO,[1,2] Z.C. ZHU,[1] X.D. WANG,[1] Y.M. SONG[1]

[1]College of Nuclear Science and Technology, University of South China, 421001 Hengyang, China

[2]Extreme Light Infrastructure-Nuclear Physics, "Horia Hulubei" National Institute for Physics and Nuclear Engineering (IFIN-HH), 30 Reactorului, 077125 Bucharest-Magurele, Romania

**Abstract**

Relativistic electrons, accelerated by the laser ponderomotive force, can be focused onto a high-$Z$ convertor to generate high-brightness beams of $\gamma$-rays, which in turn can be used to induce photonuclear reactions. In this work, the possibility of photo-transmutation of long-lived radionuclide $^{135}$Cs by laser-plasma driven electron source has been demonstrated through Geant4 simulations. High energy electron generation, bremsstrahlung and photonuclear reaction have been observed at four different laser intensities of $10^{20}$ W/cm$^2$, $5 \times 10^{20}$ W/cm$^2$, $10^{21}$ W/cm$^2$ and $5 \times 10^{21}$ W/cm$^2$, respectively. It was shown that the laser intensity and the target geometry have strong effect on the transmutation reaction yield. At different laser intensities the recommended target sizes were found to obtain the maximum reaction yield. The remarkable feature of this work is to evaluate the optimal laser intensity to produce maximum reaction yield of $10^8$ per Joule laser pulse energy, which is $10^{21}$ W/cm$^2$. Our study suggests photo-transmutation driven by laser-based electron source as a promising approach for experimental research into transmutation reactions, with potential applications to nuclear waste management.

## 1. INTRODUCTION

Beams of electrons, positrons, protons and high energy photons could be achieved by the interaction of ultra-intense laser with a solid or gas target, which received extensive attention due to the considerable potential for utilization (Ledingham *et al*., 2003; Mangles *et al*., 2004; Schwoerer *et al*., 2006; Luo *et al*., 2013; Hanus *et al*., 2014; Luo *et al*., 2015). Recently, due to the rapid evolution of ultra-intense laser technology, electrons can be accelerated by laser ponderomotive force to energies ranging from several MeV to a few hundred MeV. By focusing these relativistic electrons onto a high-$Z$ metallic target, high energy $\gamma$-rays will be generated through bremsstrahlung process, which has a wide range of applications, such as activation (or transmutation), fission, fusion (Schwoerer *et al*., 2003; Ledingham *et al*., 2003; Galy *et al*., 2007; 2009).

Photonuclear reaction that is induced by using an ultra-intense laser was initially proposed by Shkolnikov *et al.* (1997), from which the generated bremsstrahlung $\gamma$-rays and the consequent positrons, and photoneutrons were estimated. Magill *et al.* (2003) launched a photo-transmutation experiment on the long-lived radionuclide $^{129}$I to confirm the available reaction cross-section data for $^{129}$I($\gamma$,n). Then the photo-transmutation studies on radionuclides $^{135}$Cs, $^{137}$Cs, $^{90}$Sr, $^{93}$Zr and $^{126}$Sn driven by laser-based electron-bremsstrahlung sources were investigated analytically (Takashima *et al.*, 2005; Sadighi-bonabi *et al.*, 2006; Sadighi *et al.*, 2010; Sadighi-bonabi *et al.*, 2010; Irani *et al.*, 2012). These studies suggested that the parameters of laser intensity and irradiation time were closely related to the number of photonuclear reactions, and then demonstrated the possibility for the transmutation of radioactive nuclear waste by using intense lasers.

In this work, we reported a proof-of-principle experiment on the transmutation of long-lived nuclear waste $^{135}$Cs by ultra-intense laser with an intensity of $(0.1{-}5.0) \times 10^{21}$ W/cm$^2$. The long-lived radionuclide investigated, $^{135}$Cs was chosen due to its high radiotoxicity, long half-life ($T_{1/2} = 2.3$ million years) and for the visible geologic repository impact and inventory. Accordingly, it could lead to a huge risk of leakage-dose to biosphere (Yang *et al.*, 2004). Using the photo-transmutation method, $^{135}$Cs can be transmuted into $^{134}$Cs through ($\gamma$,n) reaction or into stable nuclide $^{133}$Cs through ($\gamma$,2n) reaction. For the nuclide $^{134}$Cs, it has a short half-life of 2.07 years, and beta decays to stable nuclide $^{134}$Ba. Hence these non-/low toxic or stable product nuclides could be easily handled. We investigated, through Geant4 Monte Carlo simulations (Agostinelli *et al.*, 2003), the generation of intense bremsstrahlung $\gamma$ source driven by laser-accelerated electron beam (*e*-beam) and the consequent photo-transmutation of radionuclide $^{135}$Cs. We focused on the dependence of transmutation yield on the geometry of the converting target (CT) for bremsstrahlung generation and the adjacent transmuted target (TT), such to optimize the number of transmutation reactions. Note that although the transmutation of $^{135}$Cs by ultra-intense laser has been analytically evaluated (Takashima *et al.* 2005), the influence of the laser intensity and the target geometry on the transmutation reactions is still an interesting topic and is worth to be studied. It will be helpful for the similar photonuclear experiments performed by using high-peak power lasers.

## 2. PHOTO-TRANSMUTATION MODEL

Currently laser wakefield acceleration (LWFA) and laser ponderomotive acceleration (LPA) are considered as the main acceleration schemes for electrons (Esarey *et al.*, 2009). While LWFA can deliver high-quality relativistic ($\geq$100 MeV) *e*-beams with low (a few percent) energy spread and small (a few mrad) spatial divergence, the beam current that can be accelerated was limited to be on the order of tens pC; on the contrary, LPA can distribute relativistic *e*-beams with much higher charge (up to a few nC) (Glinec *et al*., 2005; Giulietti *et al*., 2008), which could be beneficial for the increase in the bremsstrahlung $\gamma$ flux. Also should be mentioned that the LPA *e*-beams have a wide bandwidth. However, the need for narrowing LPA *e*-beam spectrum could be eliminated since the bremsstrahlung $\gamma$ source also maintains a continuous spectrum pattern. Within these considerations, we used LPA *e*-beam as the driven source to produce bremsstrahlung $\gamma$-rays, which in turn to induce photo-transmutation by irradiating them off a cesium target.

A scheme for photo-transmutation of long-lived radionuclide $^{135}$Cs by the LPA *e*-beam is illustrated schematically in Fig. 1. A metallic tantalum target was employed as bremsstrahlung convertor. Both the convertor and irradiated cesium target were assumed as ideally cylindrical structures with flexible radii and thicknesses. Since steady progress in the generation of LPA *e*-beam was made and the dependence of its spectra and angular distribution on the laser intensity has been characterized very well, we implemented directly into the Geant4 simulations the features of the LPA *e*-beams by choosing the incident laser intensities. These intensities for state-of-art lasers are $10^{20}$ W/cm$^2$, $5 \times 10^{20}$ W/cm$^2$, $10^{21}$ W/cm$^2$ and $5 \times 10^{21}$ W/cm$^2$, corresponding to pulse energies of 0.37 J, 1.86 J, 3.72 J and 18.62 J respectively when considering the focused spot size is 3 μm (FWHM). For each laser pulse, the energy conversion efficiency from laser pulse to *e*-beam was fixed to 30%, which could be achieved by selecting reasonable acceleration length (Sentoku *et al*., 2002; Tanimoto *et al*., 2009; Chen *et al*., 2009; Hanus *et al*.,

2014). According to the specified conversion efficiency, the number of electrons was normalized by the incident laser pulse energy, such to facilitate the calculation and the following comparison given by this study.

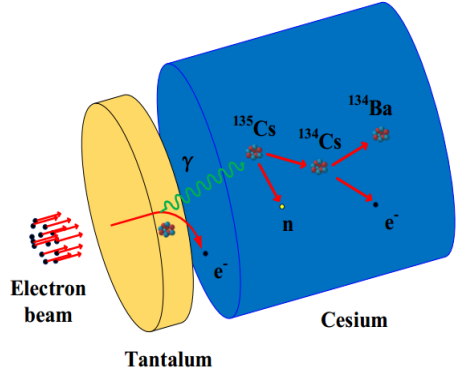

**Fig. 1** A schematic layout of transmutation of radionuclide $^{135}$Cs by laser-driven *e*-beam. Additional space has been added between the components for a better visualization of the main target structure.

In order to reduce the computing time, a total of $10^8$ electrons were used in Geant4 simulations and distributed along a Maxwellian energy distribution (Tanimoto *et al.*, 2009; Antici *et al.*, 2012)

$$f(E) = \frac{2}{\sqrt{\pi} k T_h^{3/2}} \sqrt{E_e} \exp\left(-\frac{E_e}{T_h}\right). \tag{1}$$

Here, $E_e$ is the kinematic energy of the LPA electrons, $k$ is the Boltzmann constant and $T_h$ is the electron's temperature, which was described by Wilks *et al.* (1992)

$$T_h = 0.511 \left[\sqrt{\left(1 + \frac{I \lambda_\mu^2}{1.37}\right)} - 1\right], \tag{2}$$

where $I$ is the laser intensity with a unit of W/cm$^2$ and $\lambda_\mu$ is the wavelength with a unit of μm. According to Eq. (1), we calculated the spectral distributions of the LPA *e*-beams at different laser peak intensities (see Fig. 2). It is shown that the laser peak intensity has an important impact on the accelerated *e*-beam spectrum. The higher the peak intensity of the laser, the larger the proportion of the energetic electrons. Combing with the cross-section of photonuclear reaction one could optimize the number of reactions by changing the dimensions of the convertor as well as the following cesium target, as discussed in Section 4.

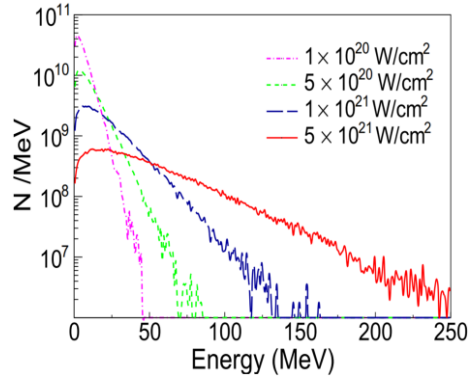

**Fig. 2** Energy spectrum of the LPA *e*-beam for the laser intensity varying from $10^{20}$ W/cm$^2$ to $5 \times 10^{21}$ W/cm$^2$.

The *e*-beam with a spot size of 3 μm (FWHM) was initialized and then oriented towards the front surface of the convertor. Since the angular distribution of the *e*-beam is closed related to the accelerated electron energy, it was characterized by a Gaussian distribution following (Moore *et al.*, 1995; Quesnel *et al.*, 1998; Debayle *et al.*, 2010)

$$\theta = \tan^{-1} \sqrt{\frac{2}{(\gamma - 1)}} \,, \tag{3}$$

where $\gamma$ is the Lorenz factor of the relativistic electrons. The transverse profile of the *e*-beam from Eq. (3) was shown in Fig. 3. Such profile was recorded at 1 cm downstream of the initial position of the *e*-beam. It is seen that the *e*-beam produced by higher laser intensity displays more collimated pattern due to the larger energy accelerated by the laser.

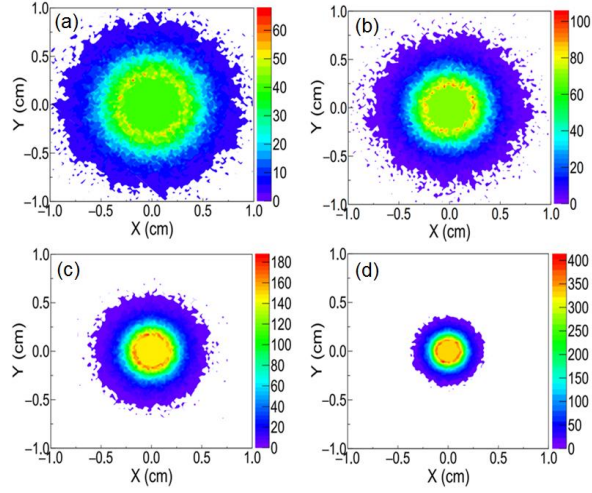

**Fig. 3** The transverse distribution of the LPA *e*-beam recorded at 1 cm downstream of the initial position of the *e*-beam for the laser intensity of $10^{20}$ W/cm$^2$ (a), $5 \times 10^{20}$ W/cm$^2$ (b), $10^{21}$ W/cm$^2$ (c), and $5 \times 10^{21}$ W/cm$^2$ (d), respectively.

## 3.   SECONDARY SOURCES DRIVEN BY LPA ELECTRON BEAM

Generally, the convolution of the bremsstrahlung spectrum and the cross-section of the photonuclear reactions is significant to reaction yield. The interaction of the LPA electrons (see Fig. 2 and Fig. 3) with the convertor was simulated at different laser intensities of $10^{20}$ W/cm$^2$, $5 \times 10^{20}$ W/cm$^2$, $10^{21}$ W/cm$^2$ and $5 \times 10^{21}$ W/cm$^2$, respectively. Consequently secondary particles such as electrons, positrons and bremsstrahlung $\gamma$-rays were generated and then diagnosed. Fig. 4 shows the bremsstrahlung spectrum, produced by the LPA *e*-beam interacting with 1.5-mm thick tantalum convertor at different laser intensities. Also shown in Fig. 4 is the total cross-sections of ($\gamma$,n) and ($\gamma$,2n) reactions on $^{135}$Cs. While the transmutation reaction has the neutron separation energy of 8 MeV, its peaked cross-section occurs at around 15 MeV. At laser intensities below $10^{21}$ W/cm$^2$, the photonuclear reaction yield caused by bremsstrahlung $\gamma$-rays increased with the enhancement of the laser intensity according to the convolution between the bremsstrahlung spectrum and the reaction cross-section, as shown in Fig. 4. However, at laser intensities above $10^{21}$ W/cm$^2$, the reaction yield can hardly increase rapidly any more.

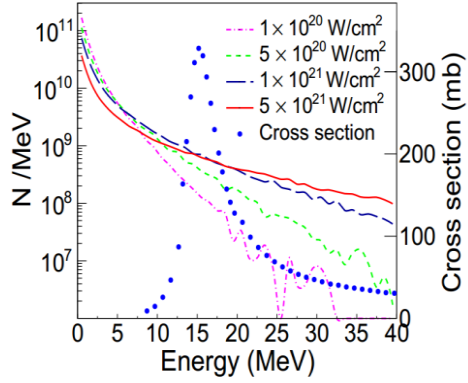

**Fig. 4** The bremsstrahlung spectrum at four different laser intensities ranging from $10^{20}$ W/cm$^2$ to $5 \times 10^{21}$ W/cm$^2$ together with the total cross-sections for $^{135}$Cs (γ,n) and $^{135}$Cs (γ,2n) reactions. The radius and the thickness of the CT are fixed to be $r_1$=2 cm and $T_1$=1.5 mm in the simulation.

Together with the bremsstrahlung γ-rays escaped from the rear face of the convertor, the emitted secondary electrons and positrons can also irradiate the TT and then produce high energy γ-rays by the processes of bremsstrahlung, which in turn trigger photonuclear reactions additionally. In order to investigate the dependence of the secondary electrons and positrons on the transmutation reaction, the electron spectrum and the positron spectrum were diagnosed and shown in Fig. 5. The target dimension was chosen as the same as in Fig. 4. It is shown that both the *e*-beam and positron beam had Maxwell-like spectral distributions. The number of high energy electrons and positrons increased with the increment of laser intensity. Similar to bremsstrahlung γ-rays, they may contribute to the transmutation yield due to the overlap of the energy spectrum with the reaction cross-sections (see Fig. 4). However, it is found that since the positrons are the minority species compared with electrons and bremsstrahlung γ-rays, their contribution could be neglected.

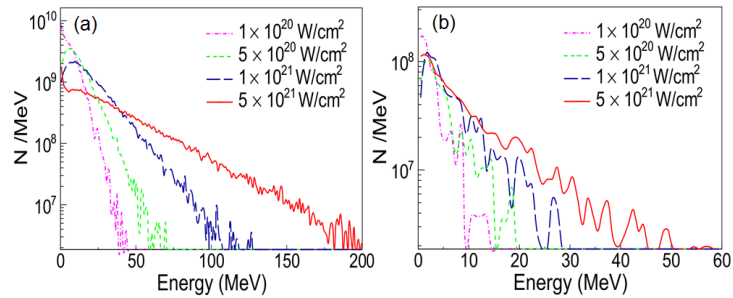

**Fig. 5** The electron spectrum (a) and positron spectrum (b) for the laser intensity varying from $10^{20}$ W/cm$^2$ to $5 \times 10^{21}$ W/cm$^2$. The target geometries were chosen as the same as in Fig. 4.

Considering that the γ-rays and electrons with energies below the neutron separation energy could not induce the photonuclear reaction, at four different laser intensities we counted the yield of electrons and γ-rays with energies above 6 MeV, as shown in Fig. 6. The yield of secondary electrons decreased as the CT thickness increased. Different from the variation trend of the electrons, the yield of γ-rays increased with the increase in the CT thickness, and then reached a saturation value when a few mm thick target is used. The consequent γ-ray yield was obtained to be the order of $10^{10}$ per Joule laser energy, which is hereafter simplified as per Joule. At different laser intensities the peaked values of $1.0 \times 10^{10}$ per Joule, $1.8 \times 10^{10}$ per Joule, $3.1 \times 10^{10}$ per Joule and $3.7 \times 10^{10}$ per Joule were obtained by using the CT thickness of 1.5 mm, 2.5 mm, 3.5 mm and 5.5 mm, respectively. These results shows that as the laser intensity used for photo-transmutation studies increased, the suggested thickness of the convertor should be increased accordingly.

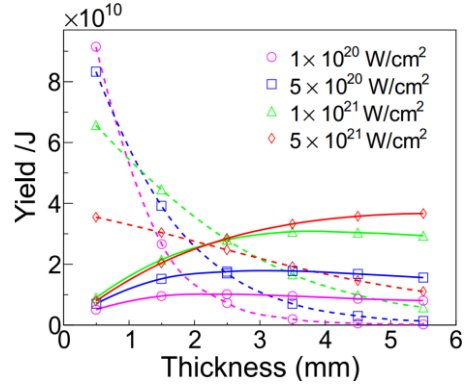

**Fig. 6** The γ-ray yield (solid lines) and the electron yield (dashed lines) per joule laser energy (/J) as a function of the CT thickness at laser intensities varying from $10^{20}$ W/cm$^2$ to $5 \times 10^{21}$ W/cm$^2$. The γ-rays and electrons with the energies over 6 MeV were taken into account due to the neutron separation energy of 8 MeV. The radius of the convertor was fixed to be $r_1$=2 cm in the simulation.

## 4. PHOTO-TRANSMUTATION OF $^{135}$CS

In this section, we discussed in detail the influence of the target parameters on the transmutation yield of $^{135}$Cs at four different laser intensities presented above. For simplicity, we only take into account three critical factors in the simulations: the thickness of the convertor, and the radius and the thickness of the transmuted target.

### 4.1 The influence of the CT thickness

Secondary sources driven by the LPA *e*-beam presented previously were employed to transmute the long-lived nuclear waste $^{135}$Cs. Since the reactions (γ,n) and (γ,2n) are the main photo-transmutation channels, only the products $^{134}$Cs and $^{133}$Cs were focused on and the consequent number of reactions were counted. Fig. 7 shows the contribution of secondary particles to transmutation reactions together with the resultant total contributions at laser intensities varying from $10^{20}$ W/cm$^2$ to $5 \times 10^{21}$ W/cm$^2$. In the simulation the radius of the CT was fixed to be 2 cm, and the radius and the thickness of the TT were chosen to be 4 cm and 3 cm, respectively. For a thin CT, the electrons made a large contribution to the transmutation reactions compared to the bremsstrahlung γ-rays. Then with the increase of the CT thickness, the contribution of the electrons reduced while that of the γ-rays increased. However, as the CT thickness increased to a certain value, the contribution of the γ-rays decreased as a result of the decrease in its yield. It is also shown in Fig. 7 that due to the contribution of electrons the thickness of the CT that led to the maximal reaction yield in total is slightly thinner than that led to the peaked γ-ray yield. It suggests that the influence of the electrons should be taken into account in order to obtain the reaction yield as reliable as possible. Such effect has not been presented in the previous literature. In addition, the contribution of the positrons was analyzed to be much smaller than 7% and consequently it was not shown in the figure.

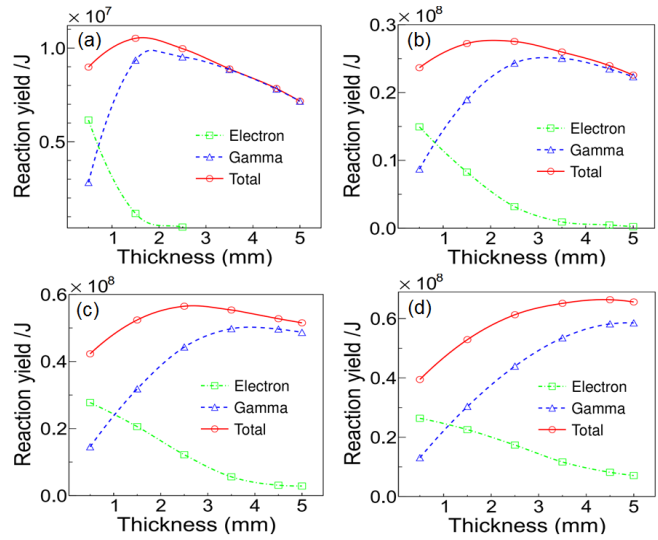

**Fig. 7** The contribution of the electrons and the bremsstrahlung $\gamma$-rays to the transmutation reaction at laser intensities of $10^{20}$ W/cm$^2$ (a), $5 \times 10^{20}$ W/cm$^2$ (b), $10^{21}$ W/cm$^2$ (c) and $5 \times 10^{21}$ W/cm$^2$ (d). The total contribution is also shown in the figure. The radius of the CT used in the simulations was $r_1$=2 cm, and the radius and the thickness of the TT were $r_2$=4 cm and $T_2$=3 cm, respectively.

Fig. 8 shows the dependence of the reaction yield on the CT thickness. As the value of the thickness of the CT is set to 0, it means the case of "without CT", from which the LPA $e$-beam irradiated on the transmuted target directly and then triggered the photonuclear reactions. One could see that with the help of the CT, the transmutation yield enhanced effectively, especially at laser intensities above $10^{21}$ W/cm$^2$. In order to obtain the maximum reaction yield, the optimized thickness for the CT are suggested to be 1.0 mm, 1.5 mm, 2.5 mm and 3.5 mm at laser intensities of $10^{20}$ W/cm$^2$, $5 \times 10^{20}$ W/cm$^2$, $10^{21}$ W/cm$^2$ and $5 \times 10^{21}$ W/cm$^2$, respectively. It is also shown in Fig. 8 that reaction yield at laser intensity of $5 \times 10^{21}$ W/cm$^2$ is relatively smaller than that at $10^{21}$ W/cm$^2$ when the CT thickness below 1.5 mm is used. This is mainly caused by the convolution of the $\gamma$ spectrum with the shape of the photonuclear cross-section, as discussed above.

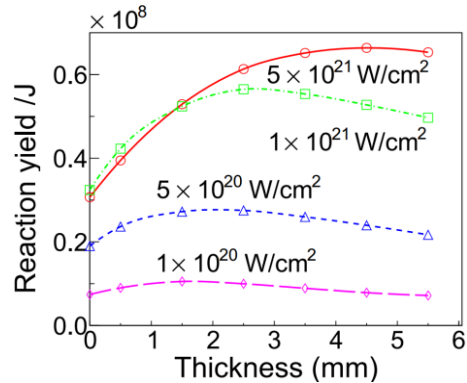

**Fig. 8** The reaction yield versus the CT thickness at laser intensities ranging from $10^{20}$ W/cm$^2$ to $5 \times 10^{21}$ W/cm$^2$. The target geometries are the same as those in Fig. 7.

### 4.2 The influence of the geometry of the transmuted target

Using the optimized thickness of the CT, the dependence of transmutation reactions on the TT geometry were investigated. The curve of the reaction yield as a function of the TT radius is investigated and is shown in Fig. 9. The reaction yield enhanced rapidly when the target radius is relatively small , e.g. ≤1.0 cm, meanwhile such enhancement ceased when the radius of the target larges 1.5 cm. Taking into account the volume of the TT , the radius of the TT is suggested to be 2 cm at four different laser intensities. In the simulation, the thicknesses of the TT was set as 3 cm.

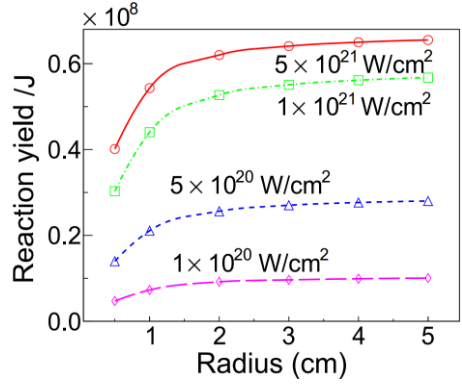

**Fig. 9** The reaction yield as a function of the TT radius at laser intensities varying from $10^{20}$ W/cm² to $5 \times 10^{21}$ W/cm². While keeping the same values for the CT radius and the TT thickness as those used in Fig. 7, i.e., $r_1 = 2$ cm, $T_2 = 3$ cm, we used the optimized convertor thickness as follows: $T_1 = 1.0$ mm, 1.5 mm, 2.5 mm and 3.5 mm at laser intensities of $10^{20}$ W/cm², $5 \times 10^{20}$ W/cm², $10^{21}$ W/cm² and $5 \times 10^{21}$ W/cm², respectively.

At four different laser intensities, the dependence of reaction yield on the TT thickness is shown in Fig. 10. For $\geq 10^{21}$ W/cm² lasers, the reaction yield increases when the thickness of the TT becomes thicker. On the contrary, at the lower laser intensities, such increase is not obvious and even ceases. The radii of the CT and the TT were fixed as 2 cm, and the optimized thicknesses of the CT were used in the simulation, as discussed above.

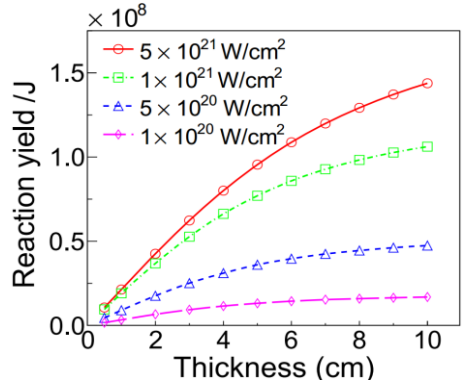

**Fig. 10** The reaction yield versus the TT thickness at different laser intensities. The optimized parameters of the targets used in the simulation are the following: $r_1 = 2$ cm, $r_2 = 2$ cm and $T_1 = 1.0$ mm, 1.5 mm, 2.5 mm and 3.5 mm at laser intensities of $10^{20}$ W/cm², $5 \times 10^{20}$ W/cm², $10^{21}$ W/cm² and $5 \times 10^{21}$ W/cm², respectively.

### 4.3 Discussion

At laser intensities ranging from $10^{20}$ W/cm² to $5 \times 10^{21}$ W/cm², the influence of the parameters for both the convertor and transmuted target has been demonstrated (see Figs. 7−10). According to these simulations, the transmutation yield of $^{135}$Cs were optimized by the target geometry parameters. To illustrate these optimizations more clearly, the reaction yields for different cases of target geometry are shown in Fig. 11. While at the laser intensities below $10^{21}$ W/cm² the reaction yield increases proportionally with its intensity, at laser intensities exceeding $10^{21}$ W/cm², the reaction yield either reaches a saturation or increases very slowly. As a result there is an optimum laser intensity at which the maximum reaction yield can be obtained. However, beyond the optimal intensity, this effect is not obvious. For the transmutation of $^{135}$Cs by ultra-intense laser, the optimum intensity has been calculated to be $10^{21}$ W/cm² with induced reaction yield of the LPA $e$-beam.

Fig. 11 also shows that the case of "with CT" is beneficial for the enhancement of transmutation yield compared to the case of "without CT". At laser intensities of $(1.0-5.0) \times 10^{20}$

W/cm$^2$ and for a 0.5-mm thick CT the reaction yield is about 1.2−1.3 times larger than that of "without CT". Then at different laser intensities of $10^{20}$ W/cm$^2$, $5 \times 10^{20}$ W/cm$^2$, $10^{21}$ W/cm$^2$ and $5 \times 10^{21}$ W/cm$^2$, the recommended CT thicknesses are obtained to be 1.0 mm, 1.5 mm, 2.5 mm and 3.5 mm, respectively. The resultant reaction yields are 1.1, 1.2, 1.3 and 1.7 times higher than those provided by the convertor with a fixed value 0.5 mm. Afterward, the TT thicknesses were optimized further to be 4 cm, 6 cm, 8 cm and 10 cm, and the resultant reaction yields are 1.2, 1.5, 1.7 and 2.2 times higher compared to that provide by a 3 cm thick TT. Finally the reaction yields obtained are $0.1 \times 10^8$ per Joule for $10^{20}$ W/cm$^2$ laser, $0.4 \times 10^8$ per Joule for $5 \times 10^{20}$ W/cm$^2$ laser, $1.0 \times 10^8$ per Joule for $10^{21}$ W/cm$^2$ laser and $1.4 \times 10^8$ per Joule for $5 \times 10^{21}$ W/cm$^2$ laser, respectively. It is obvious that laser repetition rate has direct effect on increasing the yield, but attaining higher rates requires achieving high powers with more advanced lasers. We expect that optimum ultra-intense laser ($10^{21}$ W/cm$^2$) with 1 kHz repetition rate is capable of producing about $10^{11}$ reactions, presenting the potential for affordable and tabletop lasers for nuclear transmutation studies.

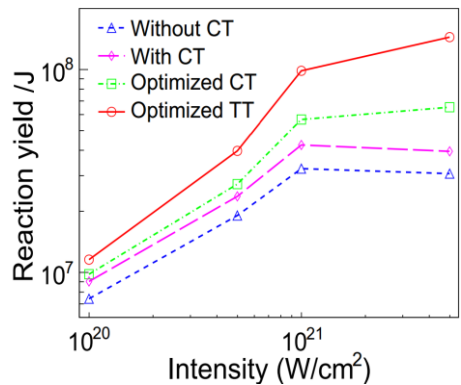

**Fig. 11** The reaction yield at four different laser intensities. For the case of "without CT", the target dimensions used in the simulations are $r_1$=2 cm, $r_2$=4 cm and $T_2$=3 cm. For the case of "with CT" are $r_1$=2 cm, $r_2$=4 cm, $T_1$=0.5 mm and $T_2$=3 cm. For the optimized CT case, are $r_1$=2 cm, $r_2$=4 cm, $T_2$=3 cm and $T_1$=1.0 mm, 1.5 mm, 2.5 mm and 3.5 mm at laser intensities of $10^{20}$ W/cm$^2$, $5 \times 10^{20}$ W/cm$^2$, $10^{21}$ W/cm$^2$ and $5 \times 10^{21}$ W/cm$^2$, respectively. For the optimized both CT and TT case, are $r_2$=2 cm, $r_2$=4 cm, $T_1$=1.0 mm, 1.5 mm, 2.5 mm and 3.5 mm and $T_2$=4 cm, 6 cm, 8 cm and 10 cm.

## 5. CONCLUSION

In this paper, the possibility of photo-transmutation of long-lived radionuclide $^{135}$Cs, into short-lived $^{134}$Cs with a half-life of 2.3 years, or into stable nuclide $^{133}$Cs has been evaluated through Monte Carlo simulations. At laser intensities of $10^{20}$ W/cm$^2$, $5 \times 10^{20}$ W/cm$^2$, $10^{21}$ W/cm$^2$ and $5 \times 10^{21}$ W/cm$^2$, by irradiating the LPA $e$-beam off metal tantalum convertor and cesium target, high energy electron generation, bremsstrahlung and photonuclear reactions have been demonstrated. It was shown that the laser intensity and the geometry of both the convertor and the cesium target have strong and direct effect on the reaction yield of $^{135}$Cs. For example, by using the optimized targets, the reaction yields obtained are $0.1 \times 10^8$ per Joule laser energy for $10^{20}$ W/cm$^2$ laser, $0.4 \times 10^8$ per Joule for $5 \times 10^{20}$ W/cm$^2$ laser, $1.0 \times 10^8$ per Joule   for $10^{21}$ W/cm$^2$ laser and $1.4 \times 10^8$ per Joule for $5 \times 10^{21}$ W/cm$^2$ laser, respectively. Finally, we evaluated the optimal laser intensity, $10^{21}$ W/cm$^2$, to produce the maximum reaction yield and beyond the optimal intensity, this effect is very slow.

Currently, a new generation of compact, ultra-intense and tabletop laser system is being developed in a number of laboratories. The laser-based transmutation systems have the being relatively compact. This paper has introduced photo-transmutation driven by laser-accelerated electron source as a promising approach for experimental research into transmutation reactions, with potential applications to nuclear waste management and medical isotope production.

**ACKNOWLEDGMENTS**

This work was supported by the National Natural Science Foundation of China (Grant Nos. 11405083 and11347028) and the Research Foundation of Education Bureau of Hunan Province, China (Grant No. 14A120). W.L. appreciates the support from the Young Talent Project of the University of South China.